\documentclass[a4paper,conference]{IEEEtran}
\usepackage{csquotes}
\usepackage[english]{babel}

\usepackage{enumitem}
\usepackage{circledsteps}

\usepackage[doi=false,isbn=false,url=false,eprint=false,backend=biber,style=ieee,maxnames=2]{biblatex}
\usepackage{amsmath,amssymb,amsfonts,mathrsfs}
\usepackage{mathtools}
\usepackage{algorithmic}
\usepackage[ruled,norelsize]{algorithm2e}
\usepackage{amsthm}

\usepackage{here}
\usepackage{graphicx}
\usepackage[caption=false]{subfig}
\usepackage{textcomp}
\usepackage[dvipsnames]{xcolor}
\usepackage{tikz}
\usetikzlibrary{arrows}
\usetikzlibrary{arrows.meta}
\usetikzlibrary{fit}
\usetikzlibrary{ext.paths.ortho}
\usetikzlibrary{shapes.geometric}
\usetikzlibrary{shadows}
\usetikzlibrary{positioning}
\usetikzlibrary{calc}
\usetikzlibrary{automata}
\usepackage{pgfplots}
\pgfplotsset{compat=newest}
\usepgfplotslibrary{groupplots}
\usetikzlibrary{patterns.meta}
\usetikzlibrary{decorations.pathreplacing}
\pgfplotsset{colormap={inferno}{
		rgb(0)=(0.001462, 0.000466, 0.013866),
		rgb(15)=(0.037668, 0.025921, 0.132232),
		rgb(30)=(0.116656, 0.047574, 0.272321),
		rgb(45)=(0.217949, 0.036615, 0.383522),
		rgb(60)=(0.316282, 0.053490, 0.425116),
		rgb(75)=(0.410113, 0.087896, 0.433098),
		rgb(90)=(0.503493, 0.121575, 0.423356),
		rgb(105)=(0.596940, 0.154848, 0.398125),
		rgb(120)=(0.688653, 0.192239, 0.357603),
		rgb(135)=(0.775059, 0.239667, 0.303526),
		rgb(150)=(0.851384, 0.302260, 0.239636),
		rgb(165)=(0.912966, 0.381636, 0.169755),
		rgb(180)=(0.956852, 0.475356, 0.094695),
		rgb(195)=(0.981895, 0.579392, 0.026250),
		rgb(210)=(0.987464, 0.690366, 0.079990),
		rgb(225)=(0.973088, 0.805409, 0.216877),
		rgb(240)=(0.947594, 0.917399, 0.410665),
		rgb(255)=(0.988362, 0.998364, 0.644924),
}}

\usetikzlibrary{external}
\usepackage{array}
\usepackage{textcomp}
\usepackage{stfloats}
\usepackage{url}
\usepackage{verbatim}
\usepackage{graphicx}
\usepackage{nicefrac}
\usepackage{nicematrix}
\usepackage{makecell}

\usepackage{ifthen} 

\usepackage{tcolorbox}
\usepackage{bm}			
\usepackage[per-mode=symbol]{siunitx}    		
\DeclareSIUnit{\dBm}{dBm}	
\DeclareSIUnit{\dBi}{dBi}	
\DeclareSIUnit{\eq}{eq}	    
\DeclareSIUnit{\sample}{S}
\DeclareSIUnit{\sqrtW}{\ensuremath{\sqrt{\text{W}}}}
\usepackage{layouts}
\usepackage{overpic}		
\usepackage{array, booktabs, xltabular, multirow} 
\usepackage{amssymb}    
\usepackage[capitalise]{cleveref}
\Crefname{figure}{Fig.}{Figs.}
\Crefname{algorithm}{Alg.}{Algs.}
\Crefname{table}{Tab.}{Tabs.}
\crefname{equation}{}{}

\usepackage[acronym]{glossaries}

\newacronym{2d}{2D}{two-dimensional}
\newacronym{3d}{3D}{three-dimensional}
\newacronym{3gpp}{3GPP}{3rd generation partnership project}
\newacronym{4g}{4G}{fourth-generation}
\newacronym{5g}{5G}{fifth-generation}
\newacronym{6g}{6G}{sixth-generation}
\newacronym{abp}{ABP}{authentication by personalisation}
\newacronym{ac}{AC}{alternating current}
\newacronym{aci}{ACI}{adjacent channel interference}
\newacronym{aclr}{ACLR}{adjacent channel leakage ratio}
\newacronym{acpr}{ACPR}{adjacent channel power ratio}
\newacronym{adc}{ADC}{analog-to-digital converter}
\newacronym{adr}{ADR}{adaptive data rate}
\newacronym{aec}{AEC}{aluminum electrolytic capacitor}
\newacronym{aes}{AES}{Advanced Encryption Standard}
\newacronym{afe}{AFE}{analog front end}
\newacronym{ag}{AG}{array gain}
\newacronym{agc}{AGC}{automatic gain controller}
\newacronym{agps}{A-GPS}{Assisted Global Positioning System} 
\newacronym{ai}{AI}{artificial intelligence}
\newacronym{aimd}{AIMD}{Additive-Increase/Multiplicative-Decrease}
\newacronym{aiot}{A-IoT}{Ambient IoT}
\newacronym{alk}{ALK}{Alkaline}
\newacronym{am}{AM}{amplitude modulation}
\newacronym{amam}{AM/AM}{amplitude modulation to amplitude modulation}
\newacronym{ambc}{AmBC}{ambient backscatter communication}
\newacronym{amc}{AMC}{automatic modulation classification}
\newacronym{amp}{AMP}{AMbient Power}
\newacronym{ampm}{AM/PM}{amplitude modulation to phase modulation}
\newacronym{aoa}{AOA}{angle-of-arrival}
\newacronym{aod}{AOD}{angle-of-departure}
\newacronym{ap}{AP}{access point}
\newacronym{api}{API}{application program interface}
\newacronym{apt}{APT}{acoustic power transfer}
\newacronym{apu}{APU}{access point unit}
\newacronym{ar}{AR}{augmented reality}
\newacronym{arima}{ARIMA}{Auto-Regressive Integrated Moving Average}
\newacronym{arp}{ARP}{Antenna Reference Point}
\newacronym{asic}{ASIC}{application specific integrated circuit}
\newacronym{ask}{ASK}{amplitude-shift keying}
\newacronym{at}{AT}{ATtention}
\newacronym{auv}{AUV}{autonomous underwater vehicle}
\newacronym{awg}{AWG}{arbitrary waveform generator}
\newacronym{awgn}{AWGN}{additive white Gaussian noise}
\newacronym{baw}{BAW}{bulk acoustic wave}
\newacronym{bb}{BB}{base-band}
\newacronym{bc}{BC}{backscatter communication}
\newacronym{bcjr}{BCJR}{Bahl-Cocke-Jelinek-Raviv}
\newacronym{bd}{BD}{backscatter device}
\newacronym{be}{BE}{Belgium}
\newacronym{ber}{BER}{bit error rate}
\newacronym{bf}{BF}{beamforming}
\newacronym{bga}{BGA}{ball grid array}
\newacronym{bjt}{BJT}{bipolar junction transistor}
\newacronym{bldc}{BLDC}{brushless DC}
\newacronym{ble}{BLE}{Bluetooth low energy}
\newacronym{blf}{BLF}{backscatter link frequency}
\newacronym{bler}{BLER}{block error rate}
\newacronym{bod}{BOD}{Brown-Out Detection}
\newacronym{bom}{BOM}{bill of materials}
\newacronym{bpsk}{BPSK}{binary phase-shift keying}
\newacronym{brp}{BRP}{beam refinement process}
\newacronym{bs}{BS}{base station}
\newacronym{bsc}{BSC}{backscatter communication}
\newacronym{bts}{BTS}{base transceiver station}
\newacronym{btwt}{bTWT}{broadcast TWT}
\newacronym{bu}{BU}{booster unit}
\newacronym{bw}{BW}{bandwidth}
\newacronym{ca}{CA}{carrier emitter}
\newacronym{cad}{CAD}{channel activity detection}
\newacronym{cars}{CARS}{calibration reference signal}
\newacronym{cbm}{CBM}{condition based maintenance}
\newacronym{cbra}{CBRA}{contention-based random access}
\newacronym{cc}{CC}{constant current}
\newacronym{cccr}{CCCR}{complementary cross-coupled rectifier}
\newacronym{ccdd}{CCDD}{cross-coupled differential drive}
\newacronym{ccdf}{CCDF}{complementary cumulative distribution function}
\newacronym{ccnn}{CCNN}{circular convolutional neural network}
\newacronym{ccs}{CCS}{correlative channel sounder}
\newacronym{cdf}{CDF}{cumulative distribution function}
\newacronym{cdma}{CDMA}{code division-multiple access}
\newacronym{cdrx}{CDRX}{connected mode DRX}
\newacronym{ce}{CE}{coverage enhancement}
\newacronym{ced}{CED}{cumulative energy density}
\newacronym{ceofdm}{CE-OFDM}{constant-envelope OFDM}
\newacronym{cept}{CEPT}{European Conference of Postal and Telecommunications Administrations}
\newacronym{cf}{CF}{cell-free}
\newacronym{cfmmimo}{CF-mMIMO}{cell-free massive MIMO}
\newacronym{cfo}{CFO}{carrier frequency offset}
\newacronym{cfra}{CFRA}{contention-free random access}
\newacronym{cir}{CIR}{channel impulse response}
\newacronym{cla}{CLA}{closed-loop approach}
\newacronym{clk}{CLK}{clock}
\newacronym{cmle}{CMLE}{concentrated maximum likelihood estimation}
\newacronym{cmos}{CMOS}{complementary metal oxide semiconductor}
\newacronym{cn}{CN}{Core network}
\newacronym{cnn}{CNN}{convolutional neural network}
\newacronym{co}{CO}{continuous operation}
\newacronym{cordic}{CORDIC}{coordinate rotation digital computer}
\newacronym{cots}{COTS}{commercial off-the-shelf}
\newacronym{cp}{CP}{cyclic prefix}
\newacronym{cpe}{CPE}{common phase error}
\newacronym{cpfsk}{CPFSK}{continuous phase frequency shift keying}
\newacronym{cpm}{CPM}{continuous phase modulation}
\newacronym{cpo}{CPO}{carrier phase offset}
\newacronym{cpt}{CPT}{capacitive power transfer}
\newacronym{cpw}{CPW}{coplanar waveguide}
\newacronym{cqi}{CQI}{channel quality indicator}
\newacronym{cr}{CR}{coding rate}
\newacronym{crc}{CRC}{cyclic redundancy check}
\newacronym{crlb}{CRLB}{Cram\'er-Rao lower bound}
\newacronym{crs}{CRS}{cell reference signal}
\newacronym{crtwt}{C-RTWT}{coordinated restricted TWT}
\newacronym{cs}{CS}{compressed sensing}
\newacronym{cse}{CSE}{channel state estimation}
\newacronym{csi}{CSI}{channel state information}
\newacronym{csp}{CSP}{contact service point}
\newacronym{css}{CSS}{chirp spread spectrum}
\newacronym{cu}{CU}{central unit}
\newacronym{cv}{CV}{constant voltage}
\newacronym{cw}{CW}{continuous wave}
\newacronym{d2d}{D2D}{device-to-device}
\newacronym{d2r}{D2R}{device-to-reader}
\newacronym{dab}{DAB}{Digital Audio Broadcasting}
\newacronym{dac}{DAC}{digital-to-analog converter}
\newacronym{daq}{DAQ}{data acquisition system}
\newacronym{das}{DAS}{distributed antenna systems}
\newacronym{db}{DB}{duty-cycled barebone}
\newacronym{dbpsk}{DBPSK}{Differential Binary Phase Shift Keying}
\newacronym{dc}{DC}{direct current}
\newacronym{dcc}{DCC}{dynamic cooperation clustering}
\newacronym{ddc}{DDC}{digital down conversion}
\newacronym{de}{DE}{drain efficiency}
\newacronym{dect}{DECT}{Digital Cordless Telecommunications}
\newacronym{dft}{DFT}{discrete Fourier transform}
\newacronym{dftsofdm}{DFT-s-OFDM}{discrete Fourier Transform spread OFDM}
\newacronym{dftsofdmfdss}{DFT-s-OFDM-FDSS}{DFT-s-OFDM with frequency-domain spectral shaping}
\newacronym{dftsofdmfdssse}{DFT-s-OFDM-FDSS-SE}{DFT-s-OFDM with FDSS and spectral extension}
\newacronym{dl}{DL}{downlink}
\newacronym{dlc}{DLC}{Distributed Laser Charging}
\newacronym{dli}{DLI}{direct link interference}
\newacronym{dlt}{DLT}{Distributed Ledger Technology}
\newacronym{dm}{DM}{diffuse multipath}
\newacronym{dma}{DMA}{Direct Memory Access}
\newacronym{dmac}{DMAC}{Direct Memory Access Controller}
\newacronym{dmc}{DMC}{diffuse multipath component}
\newacronym{dmimo}{D-MIMO}{distributed MIMO}
\newacronym{dnn}{DNN}{deep neural network}
\newacronym{doa}{DOA}{direction-of-arrival}
\newacronym{dp}{DP}{duty-cycled persistent}
\newacronym{dpd}{DPD}{digital pre-distortion}
\newacronym{dpdk}{DPDK}{Data Plane Development Kit}
\newacronym{dram}{DRAM}{dynamic random-access memory}
\newacronym{drcs}{$\Delta$RCS}{differential-radar cross section}
\newacronym{drx}{DRX}{Discontinuous Reception Mode}
\newacronym{dsb}{DSB}{double-sideband}
\newacronym{dsl}{DSL}{digital subscriber line}
\newacronym{dsp}{DSP}{digital signal processing}
\newacronym{dss}{DSS}{Dataset Storage Standard}
\newacronym{duc}{DUC}{digital up-converter}
\newacronym{dut}{DUT}{device under test}
\newacronym{dvb}{DVB}{Digital Video Broadcasting}
\newacronym{de-ewma}{DE-EWMA}{Dynamic Error-Exponentially Weighted Moving Average}
\newacronym{e2e}{E2E}{end-to-end}
\newacronym{easa}{EASA}{European Union Aviation Safety Agency}
\newacronym{ebg}{EBG}{electromagnetic bandgap}
\newacronym{ebw}{EBW}{excess bandwidth}
\newacronym{ec}{EC}{European Commission}
\newacronym{ecc}{ECC}{Electronic Communications Committee}
\newacronym{ecdf}{eCDF}{empirical cumulative distribution function}
\newacronym{ecdlp}{ECDLP}{elliptic curve discrete logarithm problem}
\newacronym{ecsp}{ECSP}{edge computing service point}
\newacronym{edlc}{EDLC}{electrostatic double-layer capacitors}
\newacronym{edrx}{eDRX}{Extended Discontinuous Reception Mode}
\newacronym{ee}{EE}{energy efficiency}
\newacronym{egprs}{EGPRS}{Enhanced Data Rates for GSM Evolution}
\newacronym{eh}{EH}{energy harvesting}
\newacronym{eirp}{EIRP}{equivalent isotropically radiated power}
\newacronym{em}{EM}{electromagnetic}
\newacronym{ema}{EMA}{endpoint modulator approximation}
\newacronym{embb}{eMBB}{enhanced Mobile Broadband}
\newacronym{en}{EN}{energy neutral}
\newacronym{end}{END}{energy-neutral device}
\newacronym{enob}{ENOB}{effective number of bits}
\newacronym{eol}{EoL}{end of life}
\newacronym{ep}{EP}{energy profiler}
\newacronym{epd}{EPD}{electronic paper display}
\newacronym{epu}{EPU}{edge processing unit}
\newacronym{er}{ER}{Energy Receiver}
\newacronym{erc}{ERC}{European Radiocommunications Committee}
\newacronym{erp}{ERP}{effective radiation power}
\newacronym[plural=ESCs,firstplural=electronic speed controllers (ESCs)]{esc}{ESC}{electronic speed control}
\newacronym{esd}{ESD}{electrostatic discharge}
\newacronym{esl}{ESL}{electronic shelf label}
\newacronym{esr}{ESR}{equivalent series resistance}
\newacronym{et}{ET}{Energy Transmitter}
\newacronym{etsi}{ETSI}{European Telecommunications Standards Institute}
\newacronym{evd}{EVD}{eigenvalue decomposition}
\newacronym{evm}{EVM}{Error Vector Magnitude}
\newacronym{ewlb}{eWLB}{Embedded Wafer Level Ball Grid Array}
\newacronym{ewma}{EWMA}{Exponentially Weighted Moving Average}
\newacronym{eipm}{EIPM}{Energy Intelligence Platform Module}
\newacronym{fa}{FA}{federation anchor}
\newacronym{fair}{FAIR}{Findability, Accessibility, Interoperability, and Reuse of digital assets}
\newacronym{fc}{FC}{fusion center}
\newacronym{fcc}{FCC}{Federal Communications Commission}
\newacronym{fcf}{FCF}{frequency correlation function}
\newacronym{fd}{FD}{front-haul distance}
\newacronym{fdd}{FDD}{frequency-division duplexing}
\newacronym{fde}{FDE}{frequency domain equalizer}
\newacronym{fdm}{FDM}{frequency-division multiplexing}
\newacronym{fdma}{FDMA}{frequency division multiple access}
\newacronym{fdss}{FDSS}{frequency-domain spectral shaping}
\newacronym{fem}{FEM}{finite element analysis}
\newacronym{fembb}{feMBB}{further enhanced mobile broadband}
\newacronym{fft}{FFT}{fast Fourier transform}
\newacronym{fh}{FH}{fronthaul}
\newacronym{fhss}{FHSS}{frequency hopping spread spectrum}
\newacronym{fifo}{FIFO}{First In, First Out}
\newacronym{fim}{FIM}{Fisher information matrix}
\newacronym{fir}{FIR}{finite impulse response}
\newacronym{fits}{FITS}{Flexible Image Transport System}
\newacronym{fl}{FL}{Federated learning}
\newacronym{fom}{FoM}{figure of merit}
\newacronym{fov}{FOV}{field of view}
\newacronym{fpga}{FPGA}{field-programmable gate array}
\newacronym{fqt}{FQT}{Fully Quantized Training}
\newacronym{fr1}{FR1}{frequency range 1}
\newacronym{fr2}{FR2}{frequency range 2}
\newacronym{fram}{FRAM}{Ferroelectric Random Access Memory}
\newacronym{fsk}{FSK}{frequency shift keying}
\newacronym{fspl}{FSPL}{free space path loss}
\newacronym{fss}{FSS}{frequency selective surface}
\newacronym{gan}{GaN}{gallium nitride}
\newacronym{gb}{GB}{grant-based}
\newacronym{gdpr}{GDPR}{general data protection regulation}
\newacronym{gf}{GF}{grant-free}
\newacronym{gmsk}{GMSK}{Gaussian minimum-shift keying}
\newacronym{gnb}{gNB}{Next Generation Node B}
\newacronym{gni}{GNI}{gross national income}
\newacronym{gnn}{GNN}{graph neural network}
\newacronym{gnss}{GNSS}{global navigation satellite system}
\newacronym{gpclk}{GPCLK}{general purpose clock}
\newacronym{gpio}{GPIO}{General-Purpose Input/Output}
\newacronym{gpl}{GPL}{GNU General Public License}
\newacronym{gprs}{GPRS}{General Packet Radio Services}
\newacronym{gps}{GPS}{Global Positioning System}
\newacronym{gpu}{GPU}{graphical processing unit}
\newacronym{grc}{GRC}{GNU Radio Companion}
\newacronym{gscm}{GSCM}{geometry‐based stochastic model}
\newacronym{gsm}{GSM}{Global System for Mobile Communications}  %
\newacronym{gspm}{GSpM}{Generalized spatial modulation}
\newacronym{gwp}{GWP}{Global Warming Potential}
\newacronym{har}{HAR}{human activity recognition}
\newacronym{harq}{HARQ}{hybrid automatic repeat request}
\newacronym{hat}{HAT}{hardware attached on top}
\newacronym{hcs}{HCS}{human-centric services}
\newacronym{hdf5}{HDF5}{Hierarchical Data Format version 5}
\newacronym{hdr}{HDR}{high data rate}
\newacronym{hfss}{HFSS}{High Frequency Simulator Software}
\newacronym{hf}{HF}{high frequency}
\newacronym{hmd}{HMD}{head-mounted display}
\newacronym{hpbm}{HPBM}{half power beam width}
\newacronym{HyMPRo}{HyMPRo}{Hybrid Multi-Path Routing algorithm}
\newacronym{i.i.d.}{i.i.d.}{independent and identically distributed}
\newacronym{i2c}{I2C}{Inter-Integrated Circuit}
\newacronym{iaq}{IAQ}{Indoor Air Quality}
\newacronym{ib}{IB}{in-band}
\newacronym{ibbc}{IBBC}{inter-band beam configuration}
\newacronym{ibo}{IBO}{input back-off}
\newacronym{ic}{IC}{integrated circuit}
\newacronym{iccs}{ICCS}{Ilmsens correlative channel sounder}
\newacronym{ici}{ICI}{intercarrier interference}
\newacronym{icnirp}{ICNIRP}{International Commission on Non-Ionizing Radiation Protection}
\newacronym{id}{ID}{information decoding}
\newacronym{idft}{IDFT}{inverse discrete Fourier transform}
\newacronym{idxm}{IDXM}{index modulation}
\newacronym{ieee}{IEEE}{Institute of Electrical and Electronics Engineers}
\newacronym{if}{IF}{intermediate-frequency}
\newacronym{ifft}{IFFT}{inverse fast-Fourier-transform}
\newacronym{iid}{i.i.d.}{independently and identically distributed}
\newacronym{iis}{IIS}{integrated information system}
\newacronym{im}{IM}{intermodulation}
\newacronym{imd}{IMD}{intermodulation distortion}
\newacronym{imu}{IMU}{inertial measurement unit}
\newacronym{inh}{InH}{indoor hotspot office}
\newacronym{io}{IO}{input/output}
\newacronym{ioe}{IoE}{Internet of Everything}
\newacronym{iot}{IoT}{Internet of Things}
\newacronym{ipt}{IPT}{inductive power transfer}
\newacronym{ipy}{IPY}{Interventions per Year}
\newacronym{iq}{IQ}{in-phase and quadrature}
\newacronym{iqi}{IQI}{IQ imbalance}
\newacronym{ir}{IR}{infrared}
\newacronym{isi}{ISI}{intersymbol interference}
\newacronym{ism}{ISM}{industrial, scientific and medical}
\newacronym{isp}{ISP}{internet service provider}
\newacronym{itu}{ITU}{International Telecommunication Union}
\newacronym{itu-r}{ITU-R}{ITU – Radiocommunication Sector}
\newacronym{jesd}{JESD}{Joint Electron Devices Engineering Council}
\newacronym{jfet}{JFET}{junction field effect transistor}
\newacronym{kpi}{KPI}{key performance indicator}
\newacronym{ktofdm}{KT-DFT-s-OFDM}{known-tail-DFT-s-OFDM}
\newacronym{kvi}{KVI}{key value indicator}
\newacronym{larva}{LARVA}{LARge Virtual Array}
\newacronym{lbt}{LBT}{Listen-Before-Talk}
\newacronym{lca}{LCA}{life cycle assessment}
\newacronym{lch}{LCH}{logical channel}
\newacronym{lco}{LCO}{lithium cobalt oxide}
\newacronym{ldo}{LDO}{low-dropout voltage regulator}
\newacronym{ldpc}{LDPC}{low-density parity-check}
\newacronym{ldr}{LDR}{low data rate}
\newacronym{led}{LED}{Light Emitting Diode}
\newacronym{less}{LESS}{Low Energy Scheduler Solution}
\newacronym{lev}{LEV}{light electric vehicle}
\newacronym{lfp}{LFP}{lithium iron phosphate}
\newacronym{lib}{LIB}{Lithium-Ion Battery}
\newacronym{lic}{LIC}{lithium-ion capacitor}
\newacronym{lid}{LID}{Lithium Iron Disulfide}
\newacronym{lidar}{LiDAR}{light detection and ranging}
\newacronym{liion}{Li-ion}{lithium-ion}
\newacronym{lipo}{LiPo}{lithium polymer}
\newacronym{lis}{LIS}{large intelligent surface}
\newacronym{llf}{LLF}{log-likelihood function}
\newacronym{llh}{LLH}{log-likelihood}
\newacronym{lls}{LLS}{link-level simulation}
\newacronym{lmd}{LMD}{Lithium Manganese Dioxide}
\newacronym{lmmse}{LMMSE}{least minimum mean square error}
\newacronym{lmo}{LMO}{lithium ion manganese oxide}
\newacronym{lna}{LNA}{low-noise amplifier}
\newacronym{lo}{LO}{local oscillator}
\newacronym{lora}{LoRa}{long range}
\newacronym{lorawan}{LoRaWAN}{long-range wide-area network}
\newacronym{los}{LoS}{line-of-sight}
\newacronym{lp}{LP}{linear programming}
\newacronym{lpf}{LPF}{low-pass filter}
\newacronym{lpt}{LPT}{laser power transfer}
\newacronym{lpwa}{LPWA}{Low Power Wide Area}
\newacronym{lpwan}{LPWAN}{low-power wide-area network}
\newacronym{lpwans}{LPWANs}{Low-Power Wide-Area Networks}
\newacronym{lqi}{LQI}{link quality indicator}
\newacronym{llr}{LLR}{log-likelihood ratio}
\newacronym{lrelu}{LReLU}{leaky rectified linear unit}
\newacronym{lrt}{LRT}{likelihood-ratio test}
\newacronym{ls}{LS}{least squares}
\newacronym{lsa}{LSA}{large synthetic array}
\newacronym{lsf}{LSF}{large-scale fading}
\newacronym{lsfc}{LSFC}{large-scale fading component}
\newacronym{lstm}{LSTM}{Long Short-Term Memory}
\newacronym{ltc}{LTC}{lithium thionyl chloride}
\newacronym{lte}{LTE}{Long Term Evolution}
\newacronym{lti}{LTI}{linear time-invariant}
\newacronym{lto}{LTO}{lithium titanate}
\newacronym{lusta}{LUSTA 5G }{Logistique mUltimodale Sécuritaire Téléopérée \& Autonome 5G}
\newacronym{m2m}{M2M}{machine to machine}
\newacronym{ma}{MA}{Movable Antennas}
\newacronym{mac}{MAC}{Medium Access Control}
\newacronym{mate}{MATE}{millimeter-wave MIMO testbed}
\newacronym{mc}{MC}{Monte Carlo}
\newacronym{mcl}{MCL}{Maximum Coupling Loss}
\newacronym{mcs}{MCS}{modulation and coding scheme}
\newacronym{mcu}{MCU}{microcontroller unit}
\newacronym{mec}{MEC}{multi-access edge computing}
\newacronym{mems}{MEMS}{micro-electromechanical systems}
\newacronym{mf}{MF}{matched filter}
\newacronym{milp}{MILP}{Mixed Integer Linear Programming}
\newacronym{mimo}{MIMO}{multiple-input multiple-output}
\newacronym{miso}{MISO}{multiple-input single-output}
\newacronym{ml}{ML}{mismatch loss}
\newacronym{mlp}{MLP}{multilayer perceptron}
\newacronym{mmf}{MMF}{matrix matched-filter}
\newacronym{mmic}{MMIC}{monolithic microwave integrated circuit}
\newacronym{mmimo}{mMIMO}{massive MIMO}
\newacronym{mmse}{MMSE}{minimum mean square error}
\newacronym{mmtc}{mMTC}{massive machine-typed communication}
\newacronym{mmwave}{mmWave}{millimeter wave}
\newacronym{mn}{MN}{matching network}
\newacronym{mosfet}{MOSFET}{metal-oxide semiconductor field effect transistor}
\newacronym{mpc}{MPC}{multipath component}
\newacronym{mpp}{MPP}{magnetic power profile}
\newacronym{mppt}{MPPT}{maximum power point tracking}
\newacronym{mqtt}{MQTT}{Message Queuing Telemetry Transport}
\newacronym{mr}{MR}{maximum ratio}
\newacronym{mrc}{MRC}{maximum ratio combining}
\newacronym{mrc_em}{MRC}{maximum ratio combining}
\newacronym{mrc_EM}{MRC}{Magnetic Resonance Coupling}
\newacronym{mrt}{MRT}{maximum ratio transmission}
\newacronym{mse}{MSE}{mean square error}
\newacronym{msk}{MSK}{Minimum-Shift Keying}
\newacronym{mtc}{MTC}{Machine-Type Communication}
\newacronym{multi-rat}{Multi-RAT}{multiple radio access technology}
\newacronym{multirat}{Multi-RAT}{Multiple Radio Access Technology}
\newacronym{music}{MUSIC}{MUltiple SIgnal Classification}
\newacronym{mux}{MUX}{Multiplexer}
\newacronym{nas}{NAS}{Neural Architecture Search}
\newacronym{navauwall}{NAVAUWALL}{AUtomated NAVigation in WALLonia}
\newacronym{nb}{NB}{narrowband}
\newacronym{nbiot}{NB-IoT}{narrowband IoT}
\newacronym{nca}{NCA}{nickel cobalt aluminum}
\newacronym{netcdf}{NetCDF}{Network Common Data Form}
\newacronym{nf}{NF}{noise figure}
\newacronym{nfc}{NFC}{near-field communication}
\newacronym{nfv}{NFV}{network function virtualization}
\newacronym{ngmn}{NGMN}{Next Generation Mobile Networks }
\newacronym{ni}{NI}{National Instruments}
\newacronym{nicd}{NiCd}{nikkel cadmium}
\newacronym{nimh}{NiMH}{nikkel metal hydride}
\newacronym{nlos}{NLoS}{non-line-of-sight}
\newacronym{nmc}{NMC}{nickel manganese cobalt}
\newacronym{nmos}{nMOS}{n-channel metal-oxide semiconductor}
\newacronym{nn}{NN}{neural network}
\newacronym{nnls}{NNLS}{non-negative least squares}
\newacronym{noma}{NOMA}{non-orthogonal multiple access}
\newacronym{np}{NP}{Neyman-Pearson}
\newacronym{npbch}{NPBCH}{Narrowband Physical Broadcast Channel}
\newacronym{npss}{NPSS}{Narrow Band Primay Synchronization Signal}
\newacronym{nr}{NR}{New Radio}
\newacronym{nrs}{NRS}{Narrow Band Reference Signal}
\newacronym{nsss}{NSSS}{Narrowband Secondary Synchronization Signal}
\newacronym{ntp}{NTP}{network time protocol}
\newacronym{oai}{OAI}{OpenAirInterface} 
\newacronym{obw}{OBW}{occupied bandwidth}
\newacronym{ofa}{OFA}{Once-for-All}
\newacronym{ofdm}{OFDM}{orthogonal frequency-division multiplexing}
\newacronym{ofdma}{OFDMA}{orthogonal frequency-division multiple access}
\newacronym{ofdmim}{OFDM-IM}{OFDM with index modulation}
\newacronym{olos}{OLoS}{obstructed-line-of-sight}
\newacronym{oma}{OMA}{orthogonal multiple access}
\newacronym{oob}{OOB}{out-of-band}
\newacronym{ook}{OOK}{on-off keying}
\newacronym{opbo}{OPBO}{output power backoff}
\newacronym{oran}{O-RAN}{open radio-access network}
\newacronym{os}{OS}{operating system}
\newacronym{osl}{OSL}{open-short-load}
\newacronym{ota}{OTA}{over-the-air}
\newacronym{otaa}{OTAA}{over-the-air authentication}
\newacronym{otaml}{OTA-TinyML}{Over-the-air TinyML}
\newacronym{ova}{OVA}{One-Versus-All}
\newacronym{p1}{P1}{Phase 1}
\newacronym{p2}{P2}{Phase 2}
\newacronym{p2p}{P2P}{point-to-point}
\newacronym{pa}{PA}{power amplifier}
\newacronym{pae}{PAE}{power-added efficiency}
\newacronym{pam}{PAM}{pulse amplitude modulation}
\newacronym{pana}{PanA}{Panel A}
\newacronym{panb}{PanB}{Panel B}
\newacronym{papr}{PAPR}{peak-to-average power ratio}
\newacronym{pb}{PB}{power beacon}
\newacronym{pc}{PC}{pilot count}
\newacronym{pcb}{PCB}{printed circuit board}
\newacronym{pce}{PCE}{power conversion efficiency}
\newacronym{pcg}{PCG}{power consumption gain}
\newacronym{pcie}{PCIe}{Peripheral Component Interconnect Express}
\newacronym{pcsi}{PCSI}{perfect channel state information}
\newacronym{pd}{PD}{powered device}
\newacronym{pdcch}{PDCCH}{physical downlink control channel}
\newacronym{pdcp}{PDCP}{packet data convergence protocol}
\newacronym{pdf}{PDF}{probability density function}
\newacronym{pdp}{PDP}{power delay profile}
\newacronym{pdsch}{PDSCH}{physical downlink shared channel}
\newacronym{pe}{PE}{processing element}
\newacronym{peb}{PEB}{positioning error bound}
\newacronym{per}{PER}{packet error rate}
\newacronym{pet}{PET}{privacy enhancing technology}
\newacronym{peet}{PET}{Polyethylene terephthalate}
\newacronym{pg}{PG}{path gain}
\newacronym{pgd}{PGD}{proximal gradient descent}
\newacronym{phy}{PHY}{physical}
\newacronym{pin}{PIN}{positive-intrinsic-negative}
\newacronym{pki}{PKI}{public key infrastructure}
\newacronym{pl}{PL}{path loss}
\newacronym{pla}{PLA}{physically large array}
\newacronym{pla2}{PLA}{polylactic acid}
\newacronym{pll}{PLL}{phase-locked loop}
\newacronym[plural=PMs,firstplural=person months (PMs)]{pm}{PM}{person month}
\newacronym{pmf}{PMF}{polymer microwave fiber}
\newacronym{pmic}{PMIC}{power management integrated circuit}
\newacronym{pmu}{PMU}{power management unit}
\newacronym{pn}{PN}{pseudo-noise}
\newacronym{po}{PO}{phase offset}
\newacronym{poc}{PoC}{proof of concept}
\newacronym{poe}{PoE}{power-over-Ethernet}
\newacronym{pps}{1PPS}{pulse per second}
\newacronym{pr}{PR}{phase reversal}
\newacronym{prach}{PRACH}{Physical Random Access Channel}
\newacronym{prb}{PRB}{Physical Resource Block}
\newacronym{prbs}{PRBs}{Physical Resource Blocks}
\newacronym{prs}{PRS}{Peripheral Reflex System}
\newacronym{ps}{PS}{Processing System}
\newacronym{psd}{PSD}{power spectral density}
\newacronym{pse}{PSE}{power sourcing equipment}
\newacronym{psk}{PSK}{phase shift keying}
\newacronym{psm}{PSM}{power saving mode}
\newacronym{pss}{PSS}{primary synchronisation signal}
\newacronym{ptp}{PTP}{precision-time protocol}
\newacronym{ptrs}{PTRS}{Phase-Tracking Reference Signals}
\newacronym{ptw}{PTW}{paging time window}
\newacronym{pv}{PV}{photovoltaic}
\newacronym{pw}{PW}{planar wavefront}
\newacronym{pwm}{PWM}{pulse width modulation}
\newacronym{qam}{QAM}{quadrature amplitude modulation}
\newacronym{qos}{QoS}{quality-of-service}
\newacronym{qpsk}{QPSK}{quadrature phase-shift keying}
\newacronym{qrrls}{QR-RLS}{QR decomposition based recursive least squares}
\newacronym{quadriga}{QuaDRiGa}{QUAsi Deterministic RadIo channel GenerAtor}
\newacronym{r2d}{R2D}{reader-to-device}
\newacronym{ra}{RA}{Random Access}
\newacronym{ram}{RAM}{random-access memory}
\newacronym{ran}{RAN}{radio access network}
\newacronym{rar}{RAR}{Random Access Response}
\newacronym{rat}{RAT}{radio access technology}
\newacronym{raw}{RAW}{restricted access window}
\newacronym{rb}{Rb}{Rubidium}
\newacronym{rbs}{RBS}{radio base station}
\newacronym{rbw}{RBW}{resolution bandwidth}
\newacronym{rc}{RC}{raised-cosine}
\newacronym{rcs}{RCS}{radar cross section}
\newacronym{rdl}{RDL}{redistribution layer}
\newacronym{re}{RE}{radio element}
\newacronym{red}{RED}{radio equipment directiv}
\newacronym{redcap}{RedCap}{reduced capability}
\newacronym{relu}{ReLU}{rectified linear unit}
\newacronym{rf}{RF}{radio frequency}
\newacronym{rfeh}{RFEH}{radio frequency energy harvesting}
\newacronym{rfic}{RFIC}{radio-frequency integrated circuit}
\newacronym{rfid}{RFID}{radio frequency identification}
\newacronym{rfpt}{RFPT}{radio frequency power transfer}
\newacronym{rfsoc}{RFSoC}{Radio Frequency System-on-Chip}
\newacronym{rfwpt}{RF-WPT}{radio frequency wireless power transfer}
\newacronym{rir}{RIR}{room impulse response}
\newacronym{ris}{RIS}{reflective intelligent surface}
\newacronym{rl}{RL}{reinforcement learning}
\newacronym{rlc}{RLC}{Radio Link Control}
\newacronym{rllmtc}{RLLMTC}{reliable low latency machine type communication}
\newacronym{rls}{RLS}{recursive least squares}
\newacronym{rms}{RMS}{root-mean-square}
\newacronym{rmse}{RMSE}{root-mean-square error}
\newacronym{rmt}{RMT}{random matrix theory}
\newacronym{rof}{RoF}{radio-over-fiber}
\newacronym{ros}{ROS}{robot operating system}
\newacronym{rpi}{RPi}{Raspberry Pi}
\newacronym{rrc}{RRC}{Radio Resource Connection}
\newacronym{rreq}{RREQ}{route request packet}
\newacronym{rsrp}{RSRP}{Reference Signals Received Power}
\newacronym{rsrq}{RSRQ}{Reference Signal Received Quality}
\newacronym{rss}{RSS}{received signal strength}
\newacronym{rssi}{RSSI}{received signal strength indicator}
\newacronym{rtc}{RTC}{real time clock}
\newacronym{rtf}{RTF}{reader talks first}
\newacronym{rtk}{RTK}{real time kinematics}
\newacronym{rts}{RTS}{ray tracing simulator}
\newacronym{rtwt}{rTWT}{restricted TWT}
\newacronym{ru}{RU}{Radio Unit}
\newacronym{ruc}{rUC}{representative Use Cases}
\newacronym{rv}{RV}{random variable}
\newacronym{rw}{RW}{RadioWeaves}
\newacronym{rx}{RX}{receiver}
\newacronym{rzf}{RZF}{regularized zero forcing}
\newacronym{s-parameter}{S-parameter}{scattering parameter}
\newacronym{sa}{SA}{system aspects}
\newacronym{sa5}{SA}{Stand Alone}
\newacronym{sar}{SAR}{specific absorption rate}
\newacronym{sbl}{SBL}{sparse Bayesian learning}
\newacronym{sc}{SC}{single carrier}
\newacronym{scomp}{SC}{Split Computing}
\newacronym{scfde}{SC-FDE}{single-carrier modulation with frequency-domain-equalization}
\newacronym{scfdma}{SCFDMA}{single-carrier frequency division multiple access}
\newacronym{scs}{SCS}{sub-carrier spacing}
\newacronym{sdap}{SDAP}{service data adaptation protocol}
\newacronym{sdg}{SDG}{Sustainable Development Goal}
\newacronym{sdm}{SDM}{sigma-delta modulator}
\newacronym{sdma}{SDMA}{spatial-division multiple access}
\newacronym{sdn}{SDN}{software-defined network}
\newacronym{sdof}{SDoF}{sigma-delta over fiber}
\newacronym{sdr}{SDR}{software-defined radio}
\newacronym{se}{SE}{spectral efficiency}
\newacronym{sei}{SEI}{specific emitter identification}
\newacronym{ser}{SER}{symbol-error rate}
\newacronym{sf}{SF}{spreading factor}
\newacronym{sfn}{SFN}{single frequency network}
\newacronym{sfo}{SFO}{sampling frequency offset}
\newacronym{sfp}{SFP}{small form-factor pluggable}
\newacronym{sha}{SHA}{Secure Hash Algorithm}
\newacronym{SigMF}{SigMF}{Signal Metadata Format}
\newacronym{sdo}{SDO}{Standards Development Organization}
\newacronym{simo}{SIMO}{single-input multiple-output}
\newacronym{sinr}{SINR}{signal-to-interference-plus-noise ratio}
\newacronym{siso}{SISO}{single-input single-output}
\newacronym{slam}{SLAM}{simultaneous localization and mapping}
\newacronym{slc}{SLC}{spatial leakage suppression}
\newacronym{slerp}{SLERP}{spherical linear interpolation}
\newacronym{sma}{SMA}{SubMiniature version A}
\newacronym{smc}{SMC}{specular multipath component}
\newacronym{smps}{SMPS}{switched mode power supply}
\newacronym{smt}{SMT}{Surface Mount Technology}
\newacronym{sndr}{SNDR}{signal-to-noise-and-distortion ratio}
\newacronym{snidr}{SNIDR}{signal-to-noise-and-interference-and-distortion ratio}
\newacronym{snir}{SNIR}{signal-to-interference-plus-noise ratio}
\newacronym{snr}{SNR}{signal-to-noise ratio}
\newacronym{soc}{SoC}{state of charge}
\newacronym{SoC}{SOC}{System on Chip}
\newacronym{sota}{SotA}{state of the art}
\newacronym{sp}{SP}{service point}
\newacronym{spdt}{SPDT}{single pole double throw}
\newacronym{spi}{SPI}{Serial Peripheral Interface}
\newacronym{spst}{SPST}{single pole single throw}
\newacronym{sram}{SRAM}{static random-access memory}
\newacronym{srd}{SRD}{short-range device}
\newacronym{srls}{SRLS}{standard recursive least squares}
\newacronym{srs}{SRS}{Sounding Reference Signal}
\newacronym{ssb}{SSB}{synchronisation signal block}
\newacronym{ssd}{SSD}{solid state drive}
\newacronym{ssq}{SSQ}{simulator sickness questionnaire}
\newacronym{sta}{STA}{station}
\newacronym{steam}{STEAM}{science, technology, engineering, the arts, and mathematics}
\newacronym{svd}{SVD}{singular value decomposition}
\newacronym{sw}{SW}{spherical wavefront}
\newacronym{swipt}{SWIPT}{simultaneous wireless information and power transfer}
\newacronym{synce}{SyncE}{Synchronous Ethernet}
\newacronym{ta}{TA}{timing advance}
\newacronym{tau}{TAU}{tracking area update}
\newacronym{tcer}{TCER}{transported to consumed energy ratio}
\newacronym{tcp}{TCP}{Transmission Control Protocol}
\newacronym{tcxo}{TCXO}{temperature compensated crystal oscillator}
\newacronym{tdce}{TD-CE}{time-domain compression and expansion}
\newacronym{tdd}{TDD}{time division duplexing}
\newacronym{tdma}{TDMA}{time division-multiple access}
\newacronym{tdoa}{TDOA}{time-difference-of-arrival}
\newacronym{teg}{TEG}{Thermoelectric Generator}
\newacronym{tsg}{TSG}{Technical Specification Group}
\newacronym{thz}{THz}{Terahertz}
\newacronym{tinyml}{TinyML}{Tiny Machine Learning}
\newacronym{tinytl}{TinyTL}{Tiny Transfer Learning}
\newacronym{tinyol}{TinyOL}{Tiny Online Learning}
\newacronym{tinyrl}{TinyRL}{Tiny Reinforcement Learning}
\newacronym{tl}{TL}{Transfer Learning}
\newacronym{to}{TO}{timing offset}
\newacronym{toa}{TOA}{time-of-arrival}
\newacronym{tof}{ToF}{time-of-flight}
\newacronym{tosm}{TOSM}{through-open-short-match}
\newacronym{tpms}{TPMS}{Tire-Pressure Monitoring System}
\newacronym{trl}{TRL}{technology readyness level}
\newacronym{trp}{TRP}{Transmission Reception Point}
\newacronym{teng}{TENG}{Triboelectric nanogenerators}
\newacronym{tsn}{TSN}{time-sensitive networking}
\newacronym{ttf}{TTF}{tag talks first}
\newacronym{ttff}{TTFF}{Time To First Fix}
\newacronym{tti}{TTI}{transmission time interval}
\newacronym{ttm}{TTM}{time to market}
\newacronym{ttn}{TTN}{The Things Network}
\newacronym{tx}{TX}{transmitter}
\newacronym{twt}{TWT}{Target Wake Time}
\newacronym{uart}{UART}{Universal Asynchronous Receiver/Transmitter}
\newacronym{uav}{UAV}{unmanned aerial vehicle}
\newacronym{uc}{UC}{use case}
\newacronym{ucie}{UCIe}{Universal Chiplet Interconnect Express}
\newacronym{udp}{UDP}{User Datagram Protocol}
\newacronym{ue}{UE}{user equipment}
\newacronym{ugv}{UGV}{unmanned ground vehicle}
\newacronym{uhd}{UHD}{USRP hardware driver}
\newacronym{uhf}{UHF}{ultra-high frequency}
\newacronym{ul}{UL}{uplink}
\newacronym{ula}{ULA}{uniform linear array}
\newacronym{ule}{ULE}{ultra low energy}
\newacronym{ulp}{ULP}{ultra-low power}
\newacronym{uMIMO}{$\mu$-MIMO}{ultra-massive Multiple-Input Multiple-Output}
\newacronym{ummtc}{umMTC}{ultra massive machine type communication}
\newacronym{un}{UN}{United Nations}
\newacronym{unow}{uNOW}{unified non-orthogonal waveform}
\newacronym{upa}{UPA}{uniform planar array}
\newacronym{up}{UP}{ultra passive}
\newacronym{ura}{URA}{uniform rectangular array}
\newacronym{urllc}{URLLC}{ultra-reliable low-latency communications}
\newacronym{us}{US}{uncorrelated scattering}
\newacronym{usrp}{USRP}{universal software radio peripheral}
\newacronym{uv}{UV}{unmanned vehicle}
\newacronym{uw}{UW}{unique word}
\newacronym{uwb}{UWB}{ultrawideband}
\newacronym{uwofdm}{UW-DFT-s-OFDM}{unique word DFT-s-OFDM}
\newacronym{v2v}{V2V}{vehicle-to-vehicle}
\newacronym{vco}{VCO}{voltage-controlled oscillator}
\newacronym{vep}{VEP}{virtual edge platform}
\newacronym{vlc}{VLC}{visible light communication}
\newacronym{vlp}{VLP}{visible light positioning}
\newacronym{vna}{VNA}{vector network analyzer}
\newacronym{voc}{VOC}{Voltatile Organic Compound}
\newacronym{votable}{VOTable}{Virtual Observatory Table}
\newacronym{vr}{VR}{virtual reality}
\newacronym{vuca}{VUCA}{volatile, uncertain, complex and ambiguous}
\newacronym{vswr}{VSWR}{Voltage Standing Wave Ratio}
\newacronym{wifi}{Wi-Fi}{Wireless Fidelity}
\newacronym{wirelesshart}{WirelessHART}{Wireless Highway Addressable Remote Transducer Protocol}
\newacronym{wur}{WuR}{wake-up radio}
\newacronym{wus}{WuS}{wake-up signal}
\newacronym{wb}{WB}{wideband}
\newacronym{wban}{WBAN}{wireless body area network}
\newacronym{wd}{WD}{Wireless distance}
\newacronym{wimax}{WiMAX}{Worldwide Interoperability for Microwave Access}
\newacronym{wlan}{WLAN}{wireless LAN}
\newacronym[plural=WPs,firstplural=work packages (WPs)]{wp}{WP}{work package}
\newacronym{wpc}{WPC}{wireless power consortium}
\newacronym{wpt}{WPT}{wireless power transfer}
\newacronym{wr}{WR}{White Rabbit}
\newacronym{wrsn}{WRSN}{wireless rechargeable sensor network}
\newacronym{wsn}{WSN}{Wireless Sensor Network}
\newacronym{wcma}{WCMA}{weather conditioned moving average}
\newacronym{xets}{XETS}{cross exponentially tapered slot}
\newacronym{xlmimo}{XL-MIMO}{extremely large-scale MIMO}
\newacronym{xr}{XR}{extended reality}
\newacronym{z3ro}{Z3RO}{zero third-order distortion}
\newacronym{zed}{ZED}{Zero-Energy device}
\newacronym{zc}{ZC}{Zadoff–Chu}

\newacronym{upp}{µPP}{microPowerProfiler}%
\makeglossaries%

\tikzset{starrow/.style={-Stealth[inset=0pt, length=2.5mm, angle'=15]}}

\newcommand{\bmr}[1]{\bm{\mathrm{#1}}}
\newcommand{\mr}[1]{\mathrm{#1}}
\newcommand{\tw}[1]{\texttt{#1}}

\newcommand{\trans}{^\mathsf{T}}
\newcommand{\herm}{^\mathsf{H}}

\newcommand{\tc}[2]{\textcolor{#1}{#2}}

\DeclareMathOperator*{\argmin}{arg\,min} 
\DeclareMathOperator*{\argmax}{arg\,max}
\DeclareMathOperator*{\diag}{diag}
\DeclareMathOperator*{\dtft}{DTFT}

\newlength\figureheight
\newlength\figurewidth

\AtBeginBibliography{\footnotesize}

\tikzset{ antenna/.pic = { \draw[draw=black, line width=0.25mm] (0,3mm) -- (1mm,5mm); 
\draw[draw=black, line width=0.25mm] (0,3mm) -- (-1mm,5mm); 
\draw[draw=black, line width=0.25mm] (-1mm,5mm) -- (1mm,5mm); 
\draw[draw=black, line width=0.25mm] (0,0) -- (0,3mm); 
\node (NAME) at (0,5mm)[anchor=south,inner sep=0.5mm] {\scriptsize\tw{\tikzpictext}};} }
\tikzstyle{arrow} = [-{Stealth[inset=0pt, length=2.5mm, angle'=20]}, line width=0.25mm] 
\tikzstyle{line-thin} = [line width=0.25mm]
\tikzstyle{box-thin} = [rounded corners=0.5mm, text justified, draw=black, line width=0.25mm]

\IEEEoverridecommandlockouts
\begin{document}
\title{Integrated Real-Time Testbed for Wideband RFID and Wireless Power Transfer}

\author{Lukas~D'Angelo\IEEEauthorrefmark{1},
        Daniel~Pöhl\IEEEauthorrefmark{2}, 
    Benjamin~Deutschmann\IEEEauthorrefmark{1},
    Erik~Leitinger\IEEEauthorrefmark{1},
    Klaus~Witrisal\IEEEauthorrefmark{1}\\
    \IEEEauthorblockA{\IEEEauthorrefmark{1}Graz University of Technology, Austria}
    \IEEEauthorblockA{\IEEEauthorrefmark{2}NXP Semiconductors Austria GmbH \& Co KG, Gratkorn, Austria}
\thanks{The financial support by the European Union’s Horizon Europe research and innovation program under Grant No. 101192113 is gratefully acknowledged.}
}
\maketitle

\begin{abstract}
This contribution presents an experimental integrated real-time $8\times 8$ \gls{dmimo} testbed for
wideband \gls{bsc} and \gls{wpt}. 
The testbed operates in the \SI{2.45}{\giga\hertz} band
with coherent sampling at \SI{200}{\mega\sample\per\second},
employs a backscatter link frequency of \SI{40}{\kilo\hertz}, and uses wideband 5G NR 
reference signals for excitation.
We evaluate the testbed by exploiting the estimated \gls{csi}
in two target applications: 
wireless power transfer towards the \gls{bd} and real-time positioning of a \gls{bd} in an indoor environment.
In conjunction with the baseband processing chain introduced,
the testbed requires less than \SI{2}{\milli\second} of total airtime to
excite the system and acquire the signals for subsequent
synchronization and \gls{csi} estimation on uplink \gls{bsc} signals.
With the \gls{csi}, we demonstrate effective energy harvesting gains of up to \SI{12}{\deci\bel}.
\end{abstract}

\glsresetall 

\section{Introduction}
A wide range of applications demands massive wireless sensor deployments, including inventory and asset tracking,
monitoring of industrial processes, environmental sensing,
supply chain monitoring, and end-to-end logistics.
Realizing such applications requires minimizing the per-device cost and maintenance effort to ensure long-term economic viability.
\Glspl{end} achieve this by eliminating the need for batteries and
reducing device complexity,
but their ultra-low power budgets severely constrain their connectivity.
Consequently, \gls{bsc} has emerged as a promising physical-layer 
technology for \glspl{end},
as evident from its adoption in billions of \gls{uhf} \gls{rfid} devices~\autocite{Arthaber15}.
In order to further reduce deployment costs, 
recent work has focused on integrating \glspl{end} into 
existing wireless infrastructures such as 5G and Wi-Fi~\autocite{Jantti2025,Poehl2025}.
While this integration is challenged by weak \gls{bsc} signals, strong \gls{dli},
and the lack of synchronization between \glspl{end} and the infrastructure~\autocite{DAngelo2024},
it enables the acquisition of wideband \gls{csi}, which can be exploited for device 
localization~\autocite{DAngeloCluster2026} and enhanced \gls{wpt}~\autocite{DeuMueKap:WC2026}.

In this paper, we present a \gls{dmimo} \gls{bsc} testbed with multiple
TXs and RXs to facilitate efficient \gls{wpt} towards backscattering \glspl{end} in the \gls{dl} and to exploit antenna diversity in the \gls{ul}. 
We introduce a wideband \gls{dmimo} signal model applicable to generic baseband signals,
such as high data-rate \gls{ofdm} transmissions, while explicitly accounting for synchronization impairments of the \gls{bd}. Based on the signal model, we derive an algorithm estimating the synchronization parameters and the \gls{csi} of the system. Having estimated the \gls{csi}, 
we perform
approximate conjugate \gls{bf} towards the \gls{bd} for \gls{wpt}.
Finally, we position the \gls{bd} by using the estimated \gls{csi} in~\autocite{DAngeloCluster2026}. 
\section{System and Signal Model}
\label{sec:10_signal-model}
\begin{figure}
    \centering
    \vspace*{-2mm}
    \includegraphics[]{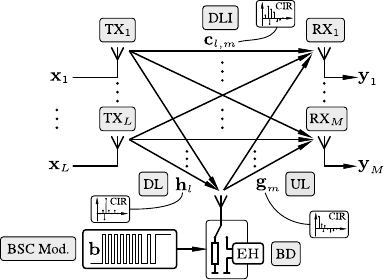}
    \vspace*{-2mm}
    \caption{Schematic overview of the discrete-time base-band equivalent \gls{dmimo} signal model.
    We have a transceiver system with $L$ TX and $M$ RX antennas, and a single-antenna energy-harvesting (EH)  \gls{bd} with a variable load impedance.
    At the RXs, we observe the sum of the direct transmission from the TXs to the respective RX, including multipath propagation,
    and the modulated signals passing through the \gls{bd}, also embedded in multipath propagation.}
    \label{fig:10_signal-model}
\end{figure}
\begin{figure}
    \centering
    \includegraphics[]{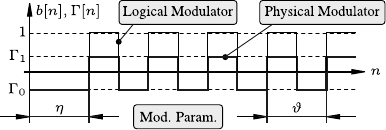}
    \vspace*{-2mm}
    \caption{Binary \gls{bsc} modulation signal as time-variant reflection coefficients. The \textit{physical modulator} 
    represents the modulation with the physical antenna mode reflection coefficients. Switching is performed according to
    a known template waveform (cf. preamble signals in~\cite{EPCc1g2}) parameterized by an unknown start time $\eta\in\mathbb{N}_0$ and symbol period $\vartheta\in\mathbb{N}$.
    The \textit{logical modulator} adheres to the same template waveform, but is forced to the domain $\{0,1\}$.
    The signals are sampled with a rate of $f_\mr{s}$, so we have $b[n] = b(nT_\mr{s})$ with a sampling period of $T_\mr{s}=f_\mr{s}^{-1}$.} 
    \label{fig:10_mod-signal}
\end{figure}

Let us consider the system and signal model in \Cref{fig:10_signal-model}.
At the \gls{bd}, information is modulated by 
altering the reflection coefficient $\Gamma$, determined
by the \textit{structural mode scattering}
and the \textit{antenna mode scattering} of the \gls{bd} \autocite{Nikitin2006}.
The former represents the reflection-independent of the actual
load impedance and the latter represents the much smaller reflection actually depending on the antenna load impedance \autocite{Guidi2014}.
The antenna mode scattering for binary \gls{bsc} is described by the physical modulation signal
$\Gamma[n] \in \{\Gamma_0,\Gamma_1\}$,
approximated with a \textit{logical modulator} $b[n] \in \{0,1\}$ as shown in \Cref{fig:10_mod-signal}. 
The samples of $b[n]$ are stacked into the vector $\bmr{b}$.
The resulting operator $\bmr{B} = \diag(\bmr{b})$ enjoys the following properties:
$\bmr{B} = \bmr{B}\herm$,
$\bmr{B} = \bmr{B}\bmr{B}$,
and there exists a \textit{complementary modulator} $\bmr{A} = \bmr{I} - \bmr{B}$,
permitting simplifications in algorithm derivation and implementation.
We use complex-valued discrete-time \gls{fir} filters to capture the propagation effects in the environment
and the hardware (e.g., antennas, cables, and \gls{bd}).
The resulting signal vector at RX$_m$ is
\begin{align}
    \bmr{y}_m = \sum_{l=1}^L\bigg(\bmr{C}_{l,m}\bmr{x}_l 
    + \bmr{G}_m\bmr{B}(\eta, \vartheta)\bmr{H}_l\bmr{x}_l\bigg)
    + \bmr{w}_m
\label{eq:10_signal-model}
\end{align}
where $\bmr{x}_l \in\mathbb{C}^{N_x\times1}$ are the samples of a wideband TX baseband signal.
The matrix 
$\bmr{C}_{l,m}\in\mathbb{C}^{(N_c + N_x-1)\times N_x}$ is the linear convolution with the direct channel from TX$_l$ to RX$_m$ $\bmr{c}_{l,m}\in\mathbb{C}^{N_c\times1}$,
$\bmr{H}_l\in\mathbb{C}^{(N_h + N_x-1)\times N_x}$ is the convolution with the $l$-th \gls{dl} channel $\bmr{h}_l\in\mathbb{C}^{N_h\times1}$,
and $\bmr{G}_m\in\mathbb{C}^{(N_g + N_h + N_x-2)\times(N_h + N_x-1)}$ is the convolution with the $m$-th \gls{ul} channel $\bmr{g}_m\in\mathbb{C}^{N_g\times 1}$.
The parameterized \gls{bsc} modulation operator is $\bmr{B}(\eta, \vartheta)=\diag(\bmr{b}(\eta, \vartheta))$ with $\bmr{b}(\eta, \vartheta)\in\{0,1\}^{(N_h + N_x-1)\times1}$, where $\eta\in\mathbb{N}_0$ is the start time of the modulation signal and $\vartheta\in\mathbb{N}$ is
the symbol period.
At each RX$_m$, we assume i.i.d. noise $\bmr{w}_m\sim\mathcal{CN}(\bmr{0}, \sigma^2\bmr{I})$.

We assume $\bmr{b}$ to be ``narrowband'' w.r.t. the transmit signals $\bmr{x}_l$  and the coherence bandwidth of the radio channels
that is large \autocite[125--143]{Molisch2011}. 
This allows us to shift the \gls{bsc} modulation to the RXs, 
so we can write the \gls{bsc} channel $\bmr{V}_{l,m}=\bmr{G}_m\bmr{H}_l$ as the convolution of \gls{dl} and \gls{ul} channels
having the cascaded \gls{cir} $\bmr{v}_{l,m}\in\mathbb{C}^{(N_h+N_g-1)\times1}$, resulting in
\begin{align}
     \bmr{y}_{m}\overset{\mathrm{EMA}}{\approx} \sum_{l=1}^L\bigg(\bmr{X}_{l}\bmr{c}_{l,m} +\widetilde{\bmr{B}}(\eta, \vartheta)\bmr{X}_{l}\bmr{v}_{l,m}\bigg) 
    + \bmr{w}_m,
\end{align}
where where $\bmr{X}_{l}\in\mathbb{C}^{(N_x+N_c-1)\times N_c}$ is the convolution with the signal $\bmr{x}_l$.
The \gls{bsc} modulator is zero-padded to compensate for the dimensionality gain of the \gls{ul} channel, 
resulting in $\widetilde{\bmr{B}}(\eta, \vartheta) = \diag\big[\bmr{b}\trans(\eta, \vartheta), \bmr{0}_{1\times (N_g\!-\!1)}\big]$.
With this approximation, we neglect the transients to the envelope $\bmr{b}(\eta, \vartheta)$ caused by the \gls{ul} \gls{cir}.
We refer to this as the \gls{ema}. For the following considerations, we assume $N_c=N_h+N_g-1$.

\section{Algorithm}
\label{sec:20_algorithm}
\subsection{Transmit Signal Separation}
First, let us reconsider the received signal in \Cref{eq:10_signal-model}.
In order to separate the different TX signals at the receivers to
get $LM$ \gls{csi} estimates $\widehat{\bmr{v}}_{l,m}\forall(l,m)$ of the \gls{bsc} channels, we employ orthogonal TX signals $\bmr{x}_l$ and analysis filters $\bmr{Q}_l\in\mathbb{C}^{(N_x+N_c-1)\times (N_x+N_c-1)}$ satisfying
\begin{align}
    \bmr{Q}_{l}\bmr{X}_{l'} = \begin{cases}
    \bmr{S}_l & l=l'\\
    \bmr{0} & l\neq l'
    \end{cases},
\end{align}
$\bmr{Q}_l\bmr{Q}_l\herm=\bmr{I}$, 
and $\bmr{Q}_l\bmr{Q}_{l'}\herm=\bmr{0}$, $\forall l\neq l'$.
With $\bmr{S}_l\in\mathbb{C}^{(N_x+N_c-1)\times N_c}$ we denote the full column rank sounding matrix.
Applying the analysis filters yields
\begin{align}
    \bmr{y}_{l,m} = \bmr{Q}_l\bmr{y}_m = \bmr{S}_{l}\bmr{c}_{l,m} +\widetilde{\bmr{B}}(\eta, \vartheta)\bmr{S}_{l}\bmr{v}_{l,m}
    + \bmr{w}_{l,m}
    \label{eq:20_separation}
\end{align}
and due to the properties of $\bmr{Q}_l$,
we get $\bmr{w}_{l,m}\sim\mathcal{CN}(\bmr{0}, \sigma^2\bmr{I})$, leaving the noise statistics and the independence assumption unaltered.
Here, the \gls{ema} is also applicable to \Cref{eq:20_separation}. A practical implementation of $\bmr{Q}_l$
is given in \Cref{eq:30_Q_mtx}.

\subsection{Synchronization and CSI Estimation}
By assuming no prior knowledge about the parameters in \Cref{eq:10_signal-model}
and by assuming stationarity of the environment during the \gls{bd}'s transmission,
we can resort to snapshot-based \gls{cmle} to obtain a practical estimator.
The strategy is to get an objective function which is concentrated
w.r.t. the linear systems $\bmr{c}_{l,m}$, $\bmr{v}_{l,m}$,
and thus depends only on the parameters $\eta$ and $\vartheta$.
Once solved for $\eta$ and $\vartheta$, we obtain closed-form solutions for 
all $\bmr{c}_{l,m}$ and $\bmr{v}_{l,m}$.

We get the joint \gls{llf} over all TXs and RXs
\begin{align}
    &\ell\big(\big\{\bmr{y}_{l,m}\big\}_{\forall l,m}; \{\bmr{c}_{l,m}, \bmr{v}_{l,m}\}_{\forall l,m}, \widetilde{\bmr{B}}(\eta, \vartheta)\big)\notag\\
    &\propto -\sum_{l,m}\big\|\bmr{y}_{l,m} - \bmr{S}_l\bmr{c}_{l,m} -\widetilde{\bmr{B}}(\eta, \vartheta)\bmr{S}_l\bmr{v}_{l,m}\big\|^2
    \label{eq:20_nice-llf}
\end{align}
with $(l,m)\in\{1,\ldots,L\}\times\{1,\ldots,M\}$.
With this we get $LM$ \gls{siso} estimates of the \gls{bsc} channels conditioned on $(\eta,\vartheta)$ and $\bmr{c}_{l,m}$, $\forall(l,m)$,
\begin{align}
    \widehat{\bmr{v}_{l',m'}} &=  \argmax_{\bmr{v}_{l',m'}}\ell\big(\big\{\bmr{y}_{l,m}\big\}_{\forall l,m}; \{\bmr{c}_{l,m}, \bmr{v}_{l,m}\}_{\forall l,m}, \widetilde{\bmr{B}}(\eta, \vartheta)\big)\notag\\
    &= \big(\widetilde{\bmr{B}}(\eta, \vartheta)\bmr{S}_{l'}\big)^+\big(\bmr{y}_{l',m'} - \bmr{S}_{l'}\bmr{c}_{l',m'}\big)
    \label{eq:20_est-bsc}
\end{align}
which can be computed independently.

Reinserting \Cref{eq:20_est-bsc} into \Cref{eq:20_nice-llf} and maximizing w.r.t. the \gls{dli} \glspl{cir} 
yields
\begin{align}
    \widehat{\bmr{c}_{l',m'}} &= \argmax_{\bmr{c}_{l',m'}}\ell\big(\big\{\bmr{y}_{l,m}\big\}_{\forall l,m}; \{\bmr{c}_{l,m}, \widehat{\bmr{v}_{l,m}}\}_{\forall l,m}, \widetilde{\bmr{B}}(\eta, \vartheta)\big)\notag\\
    &= \big(\bmr{\Pi}_{\widetilde{\bmr{B}}\bmr{S}_{l'}}^\perp\bmr{S}_{l'}\big)^+\bmr{y}_{l',m'}.
    \label{eq:20_est-dli}
\end{align}
With $\bmr{\Pi}_{\bmr{P}} \coloneqq \bmr{\Pi}(\bmr{P})= \bmr{P}\bmr{P}^+$ and $\bmr{\Pi}_{\bmr{P}}^\perp \coloneqq \bmr{I} - \bmr{\Pi}_{\bmr{P}}$
we denote the projection onto the column space and onto the space orthogonal to the column space of $\bmr{P}$, respectively.
With $\bmr{P}^+ \coloneqq (\bmr{P}\herm\bmr{P})^{-1}\bmr{P}\herm$ we denote the pseudoinverse of $\bmr{P}$.
Exploiting the properties of the logical modulator (see \cref{sec:10_signal-model}), we can simplify \Cref{eq:20_est-dli} to
\begin{align}
    \widehat{\bmr{c}_{l',m'}} &= 
    (\bmr{S}_{l'} - \widetilde{\bmr{B}}(\eta, \vartheta)\bmr{S}_{l'})^+\bmr{y}_{l',m'}.
    \label{eq:20_est-dli-nice}
\end{align}
Reinserting \Cref{eq:20_est-bsc} and \Cref{eq:20_est-dli-nice} into \Cref{eq:20_nice-llf} yields
the concentrated \gls{llf}, which is maximized w.r.t. $(\eta,\vartheta)$ to obtain synchronization
\begin{align}
    (\widehat{\eta}, \widehat{\vartheta}) &=  \argmax_{\eta,\vartheta}\ell\big(\big\{\bmr{y}_{l,m}\big\}_{\forall l,m}; \{\widehat{\bmr{c}_{l,m}}, \widehat{\bmr{v}_{l,m}}\}_{\forall l,m}, \widetilde{\bmr{B}}(\eta, \vartheta)\big)\notag\\
    &= \argmin_{\eta,\vartheta} \sum_{l,m}\big\|(\bmr{I} - \bmr{\Pi}_{\widetilde{\bmr{A}}\bmr{S}_{l}} - \bmr{\Pi}_{\widetilde{\bmr{B}}\bmr{S}_{l}})\bmr{y}_{l,m}\big\|^2,
    \label{eq:20_est-mod}
\end{align}
where $\widetilde{\bmr{A}}(\eta, \vartheta) = \bmr{I} - \widetilde{\bmr{B}}(\eta, \vartheta)$ is the \textit{complementary modulator} of
$\widetilde{\bmr{B}}(\eta, \vartheta)$. Reinserting the solution of \Cref{eq:20_est-mod} into \Cref{eq:20_est-dli-nice},
the solution of \Cref{eq:20_est-dli-nice} into \Cref{eq:20_est-bsc}, and computing \Cref{eq:20_est-bsc}, we conclude the
synchronization and \gls{csi} estimation stage.
\section{Measurement System and Evaluation}
\label{sec:30_evaluation}
\subsection{Reader Setup}
\begin{figure}[!h]
    \centering
    \includegraphics[]{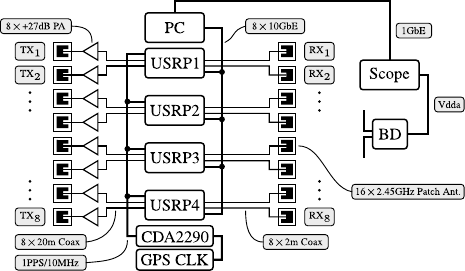}
    \vspace*{-2mm}
    \caption{Schematic overview of the measurement system, featuring $L=8$ distributed TXs (amplified by 8 \textit{Mini-Circuits ZX60-83MPR-S+} \glspl{pa}, resulting in an approx. TX power of  \SI{27}{\dBm}) and $M=8$ distributed RXs. 
    The \SI{2.45}{\giga\hertz} \SI{6}{\dBi} custom-designed directional patch antennas are connected to 4 \textit{Ettus x310} \glspl{usrp}, synchronized by
    common 1PPS and \SI{10}{\mega\hertz} reference signals, originating from a \gls{gps} clock (\textit{Agilent 58503B}) and distributed by an \textit{Ettus
    CDA2290} to allow for a sample-coherent \gls{mimo} operation. Each \gls{usrp} is linked to the PC via 2$\times$10GbE.
    A software-defined \gls{bd} from \textit{NXP} is used to demonstrate energy harvesting and \gls{bsc}-based positioning. }
    \label{fig:10_meas-schematic}
\end{figure}
\begin{figure}%
    \centering
    \centering
    \subfloat[]{{\includegraphics[width=0.532\linewidth]{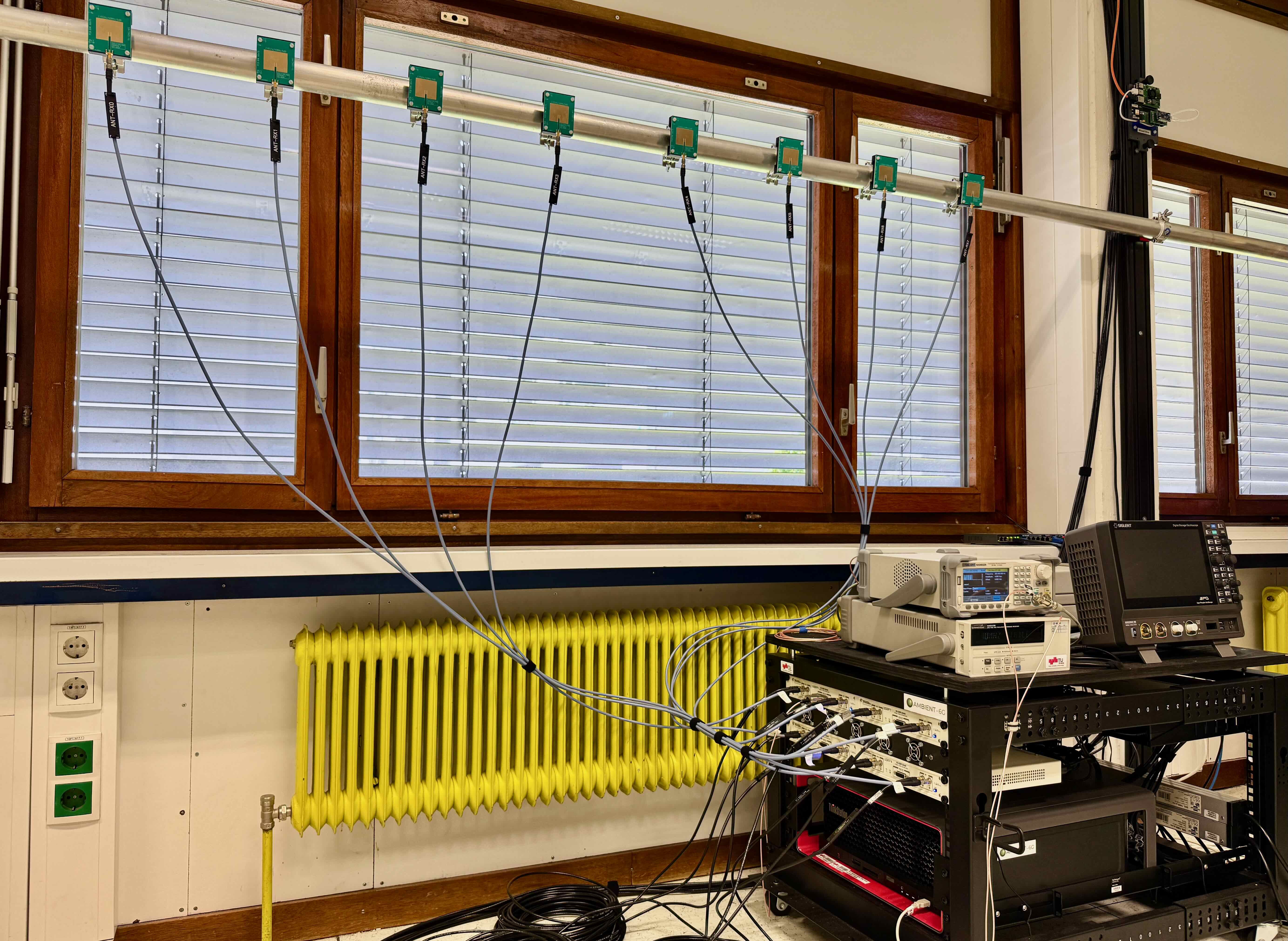} }}%
    \subfloat[]{{\includegraphics[width=0.468\linewidth]{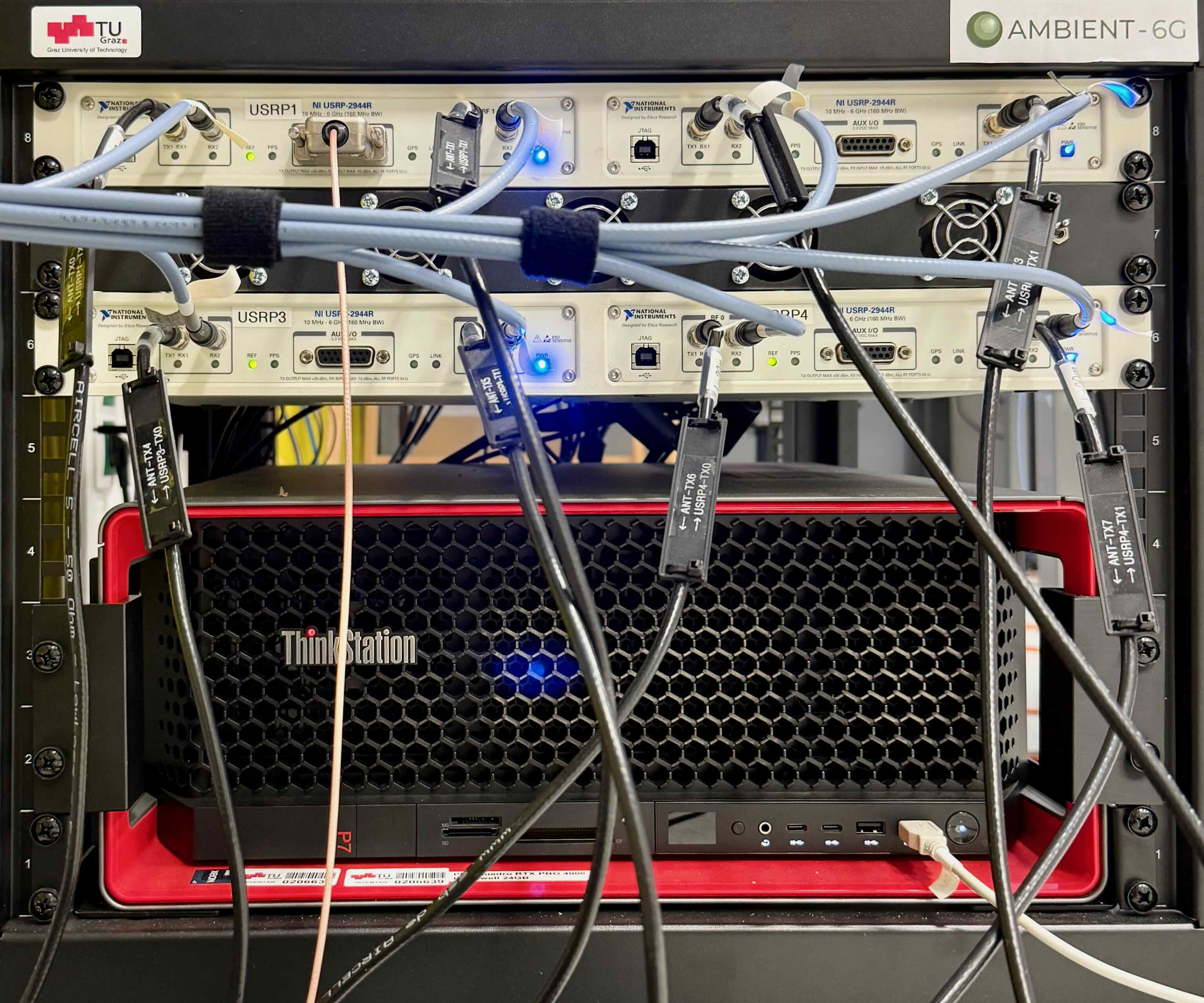} }}%
    \vspace*{-2mm}
    \caption{Central reader setup with RX antenna arrangement (a) comprising the four \glspl{usrp} and the PC (b), the CDA2290, and the GPS clock.}  %
    \label{fig:30_base-station}%
\end{figure}

\tikzexternaldisable 
\begin{figure}
    \setlength{\figurewidth}{0.85\linewidth}
    \input{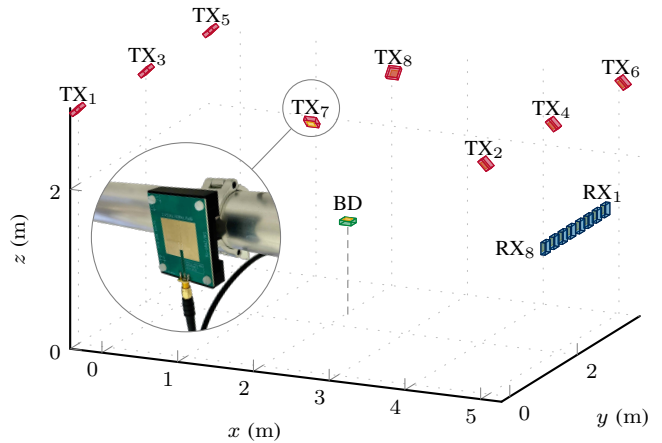}%
    \vspace{-3mm}
    \caption{Indoor measurement scenario. The 8 TX antennas are mounted on the ceiling
    of the lab with a down-tilt of $\SI{45}{\degree}$ to provide full room coverage. The
    8 RX antennas are mounted vertically at the approx. height of the \gls{bd}.}
    \label{fig:30_scenario}
\end{figure}
\tikzexternalenable

We use a \gls{sdr}-based transceiver system for the RXs and TXs, in combination with a PC
for baseband generation and processing.
This architecture enables flexible real-time implementation of algorithms, prefiltering schemes, and different 
waveforms for \gls{dl} \gls{bsc}. A schematic overview of the system is given in \Cref{fig:10_meas-schematic},
\Cref{fig:30_base-station} shows the central reader setup, and \Cref{fig:30_scenario} shows the
arrangement of the TX and RX antennas in the lab. The transceiver system operates at a carrier frequency $f_\mr{c}=\SI{2.45}{\giga\hertz}$ with a sample rate\footnote{The analog \gls{bw} of the system is mainly limited by the patch antennas,
having a \gls{bw} (return loss $<\SI{-3}{\deci\bel}$) of approx. \SI{120}{\mega\hertz}, and by the \gls{bw} of the \glspl{usrp}' \gls{rf} front-ends of approx. \SI{160}{\mega\hertz}.} of $f_\mr{s}=\SI{200}{\mega\sample\per\second}$.
A custom Python software framework using a multi-threaded implementation of the
\textit{Ettus} UHD driver governs the system, where 
synchronization in \Cref{eq:20_est-mod} and \gls{bsc} \gls{csi} estimation in \Cref{eq:20_est-bsc}
is offloaded to the GPU of the PC for a substantial speedup.
For synchronization 
and \gls{bsc} \gls{csi} estimation, we excite the system with $N_r$ \gls{zc} sequences\footnote{The employed root index $q_l\in\{25, 29, 34, 35, 44, 47, 53, 57\}$ and sequence length $N_\mr{ZC}=839$ are valid parameters for a 5G \gls{nr} long preamble of the \gls{prach}~\autocite{DAHLMAN2018}.
The param. $N_r$ is chosen s.t. $\bmr{x}_l$ extends to the approx. duration of the \gls{bd}'s transmission of \SI{2}{\milli\second}.}~\cite[148]{DAHLMAN2018} $\bmr{z}_l\in\mathbb{C}^{N_\mr{ZC}\times 1}$,
resulting in $\bmr{x}_l = \bmr{1}_{N_r}\otimes \bmr{z}_l$.
A practical approximation of the analysis filter is
\begin{align}
    \bmr{Q}\herm_l = \mr{toep}\left(\left[\bmr{z}_l\trans, \bmr{0}_{N_x + N_c - N_\mr{ZC}-1}\right]\trans\right),
    \label{eq:30_Q_mtx}
\end{align}
exhibiting low off-diagonal elements in $\bmr{Q}_l\bmr{Q}_l\herm$ and good rejection in $\bmr{Q}_l\bmr{Q}_{l'}\herm$
and $\bmr{Q}_{l}\bmr{X}_{l'}$, $l\neq l'$, due to the optimal autocorrelation and low cross-correlation properties of \gls{zc} sequences.
A computationally efficient implementation of \Cref{eq:30_Q_mtx} is achieved 
by the correlation of $\bmr{y}_m$ with the desired sequence $\bmr{z}_l$, yielding a sounding matrix $\bmr{S}_l\approx\bmr{1}_{N_r}\otimes\bmr{I}_{N_\mr{ZC}}$.

\subsection{Backscatter Device}
\begin{figure}%
    \centering
    \includegraphics[width=0.85\linewidth]{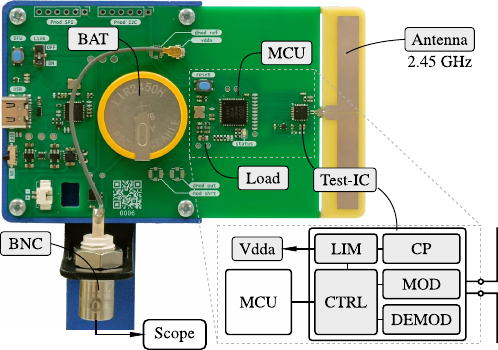}
    \vspace*{-2mm}
    \caption{Prototype of the software-defined \gls{bd}. [Detailed device description including all annotated components will follow in the extended paper.]
    } %
    \label{fig:40_bs-device}%
\end{figure}

The tag prototype in \Cref{fig:40_bs-device} implements a software-defined \gls{bd} for custom \gls{ul} transmissions. 
The \gls{rfeh} path of the Test-IC, including its charge-pump behavior, was previously characterized in~\autocite{Poehl2025}. 
Here, the same Test-IC is used as a controllable \gls{rf} front end to generate deterministic \gls{bsc} responses according to an FM0 extended preamble~\autocite{EPCc1g2} with a \gls{blf} of \SI{40}{\kilo\hertz}
for synchronization and \gls{csi} estimation. 
At the current stage, no \gls{dl} communication is decoded by the device. Instead, the integrated demodulator acts as an \gls{rf}-ON detector: after the detection of an excitation from the TXs, the \gls{mcu} triggers the programmed \gls{bsc} sequence after a predefined delay.
For the \gls{wpt} evaluation, the charge-pump output $V_{\mathrm{dda}}$ is connected to a \SI{110}{\kilo\ohm} load in parallel with the \SI{1}{\mega\ohm} oscilloscope input impedance, resulting in an effective load of approximately \SI{100}{\kilo\ohm}. 
The measured voltage across this load is used to compare the harvested power with and without \gls{bf}.

\subsection{Measurement Results}
We compare $V_{\mathrm{dda}}$ for two excitation strategies to evaluate the \gls{wpt} capability of the testbed
for the first target application. 
First, the \gls{bd} is excited without \gls{csi}-based \gls{bf} using orthogonal \gls{zc} sequences from all TXs concurrently. Second, the estimated \gls{csi} is used to prefilter the transmitted signals such that incident signals at the \gls{bd} are coherently combined.
The prefilters $\bmr{f}_l$ employed at each TX are approximate matrix-matched-filters~\cite[476]{Messerschmitt2004} $f_l[n] = \frac{1}{M}\sum\widehat{v}^*_{l,m}[-n]$, being the time-reversed~\autocite{Strohmer2004} and complex-conjugate \gls{bsc} \gls{cir}
averaged over all RXs.

The charge-pump output voltage $V_{\mathrm{dda}}$ is measured across the effective load of approx. \SI{100}{\kilo\ohm}.
After settling, the average measured voltage increases from \SI{0.323}{\volt} without \gls{bf} to \SI{1.275}{\volt} with \gls{bf}. 
Using $P_{\mathrm{harv}}=V_{\mathrm{dda}}^2/R_{\mathrm{load}}$ this corresponds to a harvested \gls{dc} power increase from \SI{1.05}{\micro\watt} to \SI{16.41}{\micro\watt}, or \SI{11.94}{\deci\bel}. 
To relate these operating points to the required incident \gls{rf} power, we use the Test-IC characterization reported in~\autocite{Poehl2025}, which was performed with the same power de-embedding procedure and a target load of \SI{100}{\kilo\ohm} configured at the microPowerProfiler~\autocite{Poehl2025}. 
At the lower voltage operating point, the estimated \gls{rf} input power is $-17.66$~dBm. At the higher voltage operating point, this increases to $-11.76$~dBm. Thus, the \gls{rf} input power increases by \SI{5.9}{\deci\bel}.
This larger \gls{dc} gain is explained by the improved \gls{pce} of the Test-IC, which increases from \SI{5.9}{\percent} to \SI{24.0}{\percent}. 
The result demonstrates that \gls{csi}-based \gls{bf} not only increases the incident \gls{rf} power at the \gls{bd}, but also shifts the energy harvester into a substantially more efficient operating region.
In \autocite{DAngeloCluster2026}, we evaluate the use of \gls{csi} for positioning, 
achieving 
an RMS error below $\SI{0.22}{\meter}$.

\tikzexternaldisable
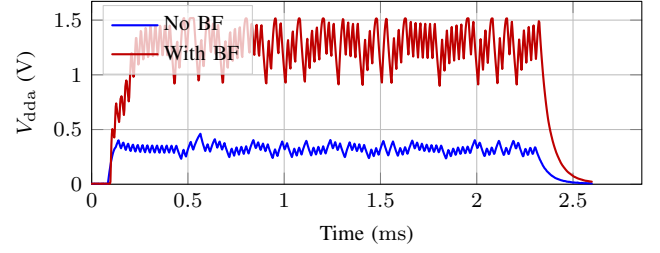
\begin{figure}
    \centering
    \begin{tikzpicture}
        \tikzset{
          every node/.style = {font=\footnotesize},
          every label/.style = {font=\footnotesize},
          every edge/.style = {font=\footnotesize},
        }
        \begin{axis}[
            height=4cm,
            width=\linewidth,
            xlabel={Time (\SI{}{\milli\second})},
            ylabel={$V_{\mathrm{dda}}$ (\SI{}{\volt})},
            xmin=0,
            ymin=0,
            grid=both,
            line join=round,
            line cap=round,
            legend style={
                at={(0.02,0.98)},
                anchor=north west,
                draw=black!20,
                fill=white,
                fill opacity=0.75,
                text opacity=1,
                font=\small
            },
            legend cell align={left},
        ]

        \addplot[
            thick,
            blue
        ] table[
            col sep=comma,
            x=time_ms,
            y=mean_vdda_no_bf
        ] {./figures/40_mean_vdda.csv};
        \addlegendentry{No \gls{bf}}

        \addplot[
            thick,
            red!75!black
        ] table[
            col sep=comma,
            x=time_ms,
            y=mean_vdda_with_bf
        ] {./figures/40_mean_vdda.csv};
        \addlegendentry{With \gls{bf}}

        \end{axis}
    \end{tikzpicture}
    \vspace*{-3mm}
    \caption{
    Measured charge-pump output voltage $V_{\mathrm{dda}}$ of the \gls{bd} averaged over five traces without and with \gls{csi}-based \gls{bf}. 
    The voltage is measured across the effective charge-pump load described in \Cref{fig:40_bs-device}. 
    The increased voltage for the \gls{bf} case indicates a significant improvement in \gls{rf} energy delivery to the \gls{bd}.
    }
    \label{fig:40_vdda_traces}
\end{figure}
\tikzexternalenable

\section{Conclusion}
\label{sec:40_conclusion}
In this work, we present an integrated real-time \gls{dmimo} \gls{bsc} testbed
for wideband \gls{rfid}, implementing synchronization, \gls{csi} estimation, \gls{bf} for \gls{wpt} and single-shot positioning of a \gls{bd}
in a realistic indoor scenario. 
The \gls{wpt} evaluation shows that \gls{csi}-based \gls{bf} increases the estimated \gls{rf} input power at the \gls{bd} by approximately \SI{5.9}{\deci\bel}, while the harvested \gls{dc} power increases by approximately \SI{12.0}{\deci\bel} due to the improved \gls{pce} of the Test-IC at the higher operating point.
These results demonstrate the validity of the proposed signal model and processing chain.

With the integrated testbed, we aim to practically validate future developments related to \gls{bsc},
including initial access schemes for \gls{wpt}, \gls{dl} and \gls{ul} communication, enhanced synchronization and \gls{csi} estimation algorithms,
and novel positioning concepts.
\printbibliography
\end{document}